# Microwave Power-to-Frequency Transduction in a Magnetically Initiated Rydberg Dissipative Time Crystal

D. Arumugam
*Jet Propulsion Laboratory, California Institute of Technology*
(Electronic mail: darmindra.d.arumugam@jpl.nasa.gov.)

Microwave power-to-frequency transduction is demonstrated using the emergent oscillation frequency of a magnetically initiated Rydberg dissipative time crystal (DTC). In a room-temperature $^{87}$Rb vapor, a ∼14 G magnetic field establishes a self-sustained DTC near 21 kHz, while a near-resonant 8.325 GHz field coupling the $63D_{3/2}\rightarrow 64P_{3/2}$ manifold continuously shifts the autonomous oscillation frequency. The resulting field-to-frequency transfer function is strongly nonlinear, with a total fundamental-frequency excursion of ∼6.7 kHz and peak responsivity approaching 1 kHz/(mV/cm). The nonlinear transition is substantially steeper than a simple quadratic saturation response and is captured by a self-consistent mean-field picture in which microwave-driven Rydberg-population redistribution modifies the collective resonance condition. Higher DTC harmonics preserve the same normalized transfer function while exhibiting approximately linear growth of absolute responsivity with harmonic order. These results establish an intrinsic many-body atomic route to RF amplitude-to-frequency conversion, enabling continuous frequency-domain sensing through an emergent atomic frequency without an externally engineered RF feedback oscillator, repeated optical spectral scans, or resolved Autler–Townes splitting.

Rydberg atoms provide sensitive and atomically referenced RF and microwave electrometry through their large transition dipole moments and well-defined energy structure. Conventional approaches infer RF amplitude from optical observables, including EIT/Autler–Townes spectral splitting and optically detected heterodyne signals [1–3]. Transduction of a measurand into frequency is widely used in precision sensing because frequency can be measured with high accuracy and is less directly dependent on absolute signal amplitude [4,5]. Recently, microwave power-to-frequency transduction was demonstrated in a Rydberg receiver using injection pulling of a self-sustained oscillator [6]. Although this approach provides large power-to-frequency responsivity, it relies on an externally engineered RF feedback oscillator and injection-pulling dynamics. It remains to be shown whether RF power-to-frequency transduction can be realized intrinsically through atomic dynamics, without an external RF oscillator or feedback loop.

Strongly interacting Rydberg ensembles provide a natural route to intrinsic nonlinear transduction. Rydberg interactions can produce population-dependent level shifts, optical bistability, and nonequilibrium phase transitions [7], while operation near collective critical points can strongly enhance microwave response [8]. Resonant microwave coupling between Rydberg states can reshape this nonlinear response [9], enabling RF sensitivity governed by many-body feedback rather than single-atom susceptibility alone.

Dissipative Rydberg time crystals (DTCs) extend this collective physics into the temporal domain, producing self-sustained oscillations with an emergent frequency [10]. Recent experiments have demonstrated RF-induced bifurcation between DTC phases [11], multiple harmonic and subharmonic time-crystalline states [12], electric-field sensing through DTC dynamics [13], Stark-induced frequency shifts and modulation [14], injection pulling and locking [15], and RF-controlled frequency tuning, intermodulation, and comb formation [16]. These studies establish strong external control of DTC dynamics and field sensing, but leave the DTC frequency itself largely unexplored as a quantitative RF sensing observable.

Here, we demonstrate resonant microwave power-to-frequency transduction in a magnetically initiated Rydberg DTC. Resonant coupling between neighboring Rydberg states shifts the autonomous DTC frequency, enabling direct mapping of microwave field amplitude onto a frequency-domain observable. We quantify the field amplitude-to-frequency responsivity and show that the nonlinear response is consistent with collective feedback arising from microwave-driven Rydberg-state redistribution. These results establish an intrinsic atomic RF amplitude-to-frequency transduction mechanism based on the emergent DTC oscillation frequency.

## I. MAGNETICALLY INITIATED DTC AND MICROWAVE COUPLING

The experimental configuration and level structure are shown in Fig. 1(a). Counter-propagating 780- and 480-nm fields address the $5S_{1/2}, F=2\rightarrow 5P_{3/2}, F'=3\rightarrow 63D_{3/2}$ ladder in a room-temperature $^{87}\mathrm{Rb}$ vapor cell (56 mm long, 25 mm diameter). The probe and coupling beams have $\sim$ 1-mm $1/e^2$ diameters with incident powers of $312\ \mu\mathrm{W}$ and $768\ \mathrm{mW}$, respectively. The probe is referenced to the $F=2\rightarrow F'=3$ transition by Doppler-free saturated-absorption spectroscopy, while the 480-nm field is generated by resonant doubling of a tunable 960-nm diode laser. Both optical fields have instantaneous linewidths below 90 kHz. Transmission of the probe through the vapor provides the optical readout of both the EIT response and the emergent temporal dynamics.

A static magnetic field applied along the optical axis lifts the Zeeman degeneracy and establishes the nonlinear regime in which persistent DTC oscillations emerge. At the operating point used here, $B \approx 14\,\mathrm{G}$, the dominant DTC frequency lies near 21 kHz. The microwave drive, applied near 8.325 GHz, couples $63D_{3/2}$ to $64P_{3/2}$ and therefore perturbs the internal Rydberg manifold rather than directly forcing the much lower DTC oscillation frequency. This distinction is central to the present transduction mechanism. As seen in Fig. 1(b), the magnetic field substantially restructures the bare EIT resonance at ∼14 G, and resonant microwave coupling produces an additional modification of the dressed optical response. The corresponding time-domain and spectral data in Figs. 1(c) and 1(d) show that the self-sustained oscillatory state persists under RF excitation while its dominant frequency shifts with applied microwave field in a repeatable manner. The bare EIT response is comparatively smooth in Fig. 1(c), whereas magnetic-field initiation produces pronounced temporal modulation that appears as strong structure across the coupler scan, reflecting the onset of DTC dynamics. Resonant microwave excitation modifies this structure and shifts the dominant DTC frequency, with repeated RF-power sweeps in Fig. 1(d) producing smooth and reversible spectral trajectories.

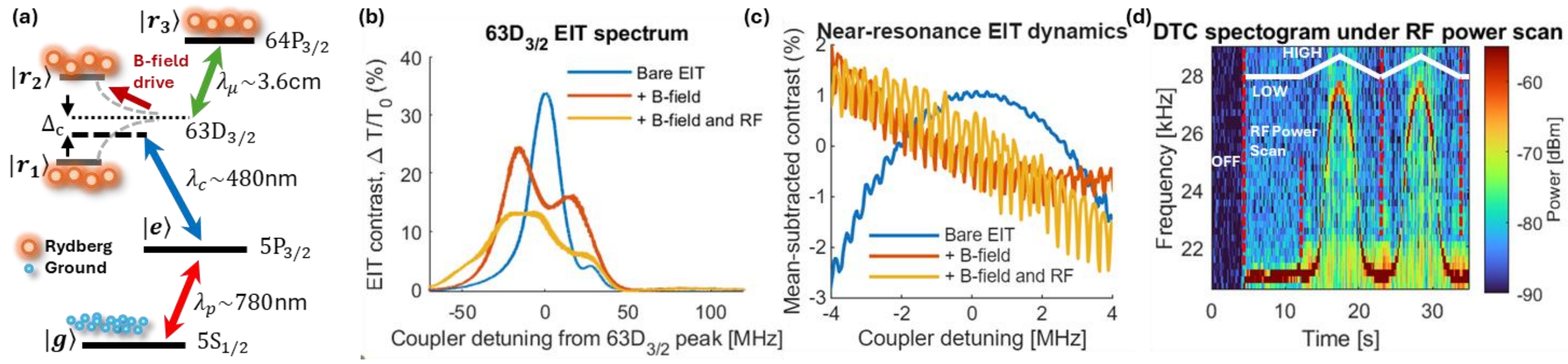

**Figure 1. Magnetically initiated Rydberg-DTC microwave power-to-frequency transduction.** (a) Concept and energy-level schematic. Rb atoms are optically excited to the $63D_{3/2}$ Rydberg state, where a static magnetic field of ∼14 G reproducibly initiates a dissipative time-crystal (DTC) oscillation. An RF field near the $63D_{3/2} \rightarrow 64P_{3/2}$ transition dresses the Rydberg manifold and changes the driven-dissipative conditions that set the DTC limit-cycle frequency. This converts microwave power into a directly measurable DTC frequency shift, providing a fixed-point optical readout that does not require resolving Autler–Townes splitting. (b) $63D_{3/2}$ EIT spectra for bare EIT, magnetic-field operation, and combined magnetic-field/RF operation. The RF-on trace is at 8.325 GHz and with a field of ∼1.2 V/m at the vapor cell, illustrating that the microwave field modifies the Rydberg-EIT response while the magnetic field establishes the self-oscillating operating point. (c) Mean-subtracted near-resonance EIT dynamics showing that the applied magnetic field turns on a stable DTC oscillation, and RF coupling dresses the ensemble and modifies the frequency. (d) DTC spectrogram during a triangular RF power scan. The initial interval corresponds to B = 0 (OFF), where no stable DTC oscillation is observed. After the magnetic field is applied, the RF field is swept over an estimated range of ∼0.22 to 2.16 V/m at the cell, as indicated by the white trace. The dominant DTC spectral component shifts upward with increasing RF power, demonstrating microwave power-to-frequency transduction.

**Figure 2. Magnetic tuning of DTC oscillations without RF coupling.** EIT spectrogram acquired with the lasers locked and no applied RF field during a triangular Helmholtz-coil field sweep, with the inferred magnetic field shown below. The DTC fundamental follows the applied field reproducibly, reaching $\sim 21\mathrm{kHz}$ near $B \approx 14\mathrm{G}$, with corresponding higher-order harmonics. The sweep identifies a magnetic bias point that initiates stable DTC oscillations while tuning the fundamental away from fixed technical tones, including the $\sim 20\mathrm{kHz}$ PID-modulation artifact.

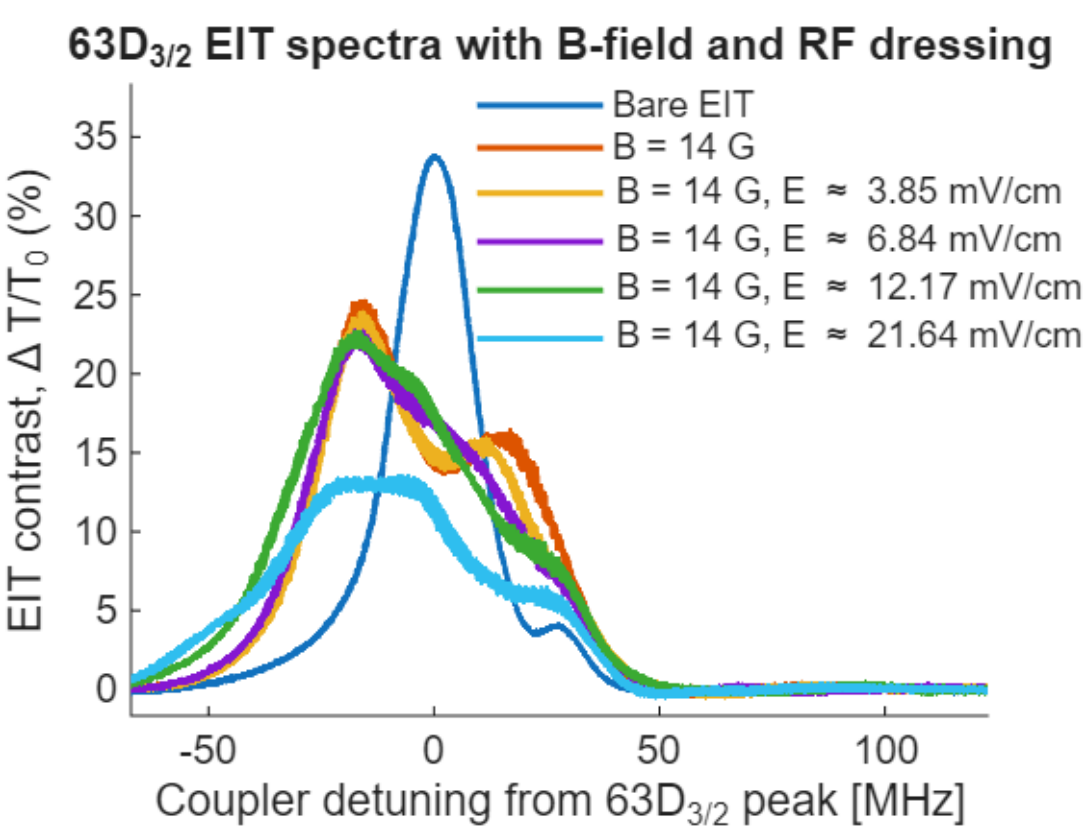

**Figure 3. EIT response under magnetic-field and RF dressing.** Coupler-detuning spectra for bare $63D_{3/2}$ EIT, magnetic-field operation at $B \approx 14\mathrm{G}$, and combined magnetic-field/RF dressing at increasing estimated RF fields. The applied magnetic field redistributes the near-resonant EIT response from a narrow, symmetric bare-EIT feature into a broader, multi-peaked structure, consistent with magnetic mixing of the Rydberg manifold and the onset of self-sustained DTC dynamics. RF coupling further modifies the dressed EIT lineshape and redistributes optical contrast across detuning, showing that the microwave field perturbs the magnetically induced DTC operating regime.

The magnetic-field dependence of the oscillatory state is isolated in Fig. 2. In the absence of RF coupling, the DTC fundamental varies reproducibly with magnetic field, with higher-order spectral components following the same evolution. The $B \approx 14$ G operating point was selected because it provides a stable ~ 21-kHz DTC while separating the fundamental from a fixed technical feature near 20 kHz. This establishes a well-defined magnetic operating point from which the microwave-induced frequency response can be measured without ambiguity from neighboring technical tones (>20 kHz). The coupler-detuning spectra in Fig. 3 provide complementary evidence that the microwave perturbation acts directly on the Rydberg manifold supporting the DTC. Relative to bare $63D_{3/2}$EIT, the magnetic field produces a broadened and partially resolved multi-feature spectrum associated with Zeeman splitting and redistribution among the participating Rydberg sublevels. Increasing resonant microwave field further reshapes this spectrum, demonstrating that the RF drive modifies the same dressed manifold that supports the self-oscillation. The field-dependent DTC frequency can therefore be viewed as the dynamical consequence of resonant microwave perturbation of the interacting Rydberg manifold, rather than as direct low-frequency forcing of the oscillator.

## II. NONLINEAR POWER-TO-FREQUENCY TRANSDUCTION

With the magnetic bias fixed near 14 G, resonant microwave excitation produces a continuous displacement of the DTC spectrum rather than a transition to an independently driven oscillatory state. In Fig. 4, the fundamental near 21 kHz evolves smoothly and reproducibly during repeated RF-power sweeps, with each higher-order harmonic undergoing a correspondingly larger absolute frequency excursion. The preservation of the harmonic structure throughout the sweep indicates that the microwave field primarily tunes the frequency of the existing DTC limit cycle. Moreover, the frequency evolution follows both increasing and decreasing RF excitation over successive cycles, demonstrating reversible power to DTC-frequency conversion over the investigated field range. Because the microwave drive acts near the $63D_{3/2} \rightarrow 64P_{3/2}$ transition rather than near the DTC frequency itself, this behavior is distinct from DTC injection pulling or locking in [15]: the GHz field modifies the internal Rydberg dynamics, and the resulting change appears as a shift of the emergent kHz oscillation frequency.

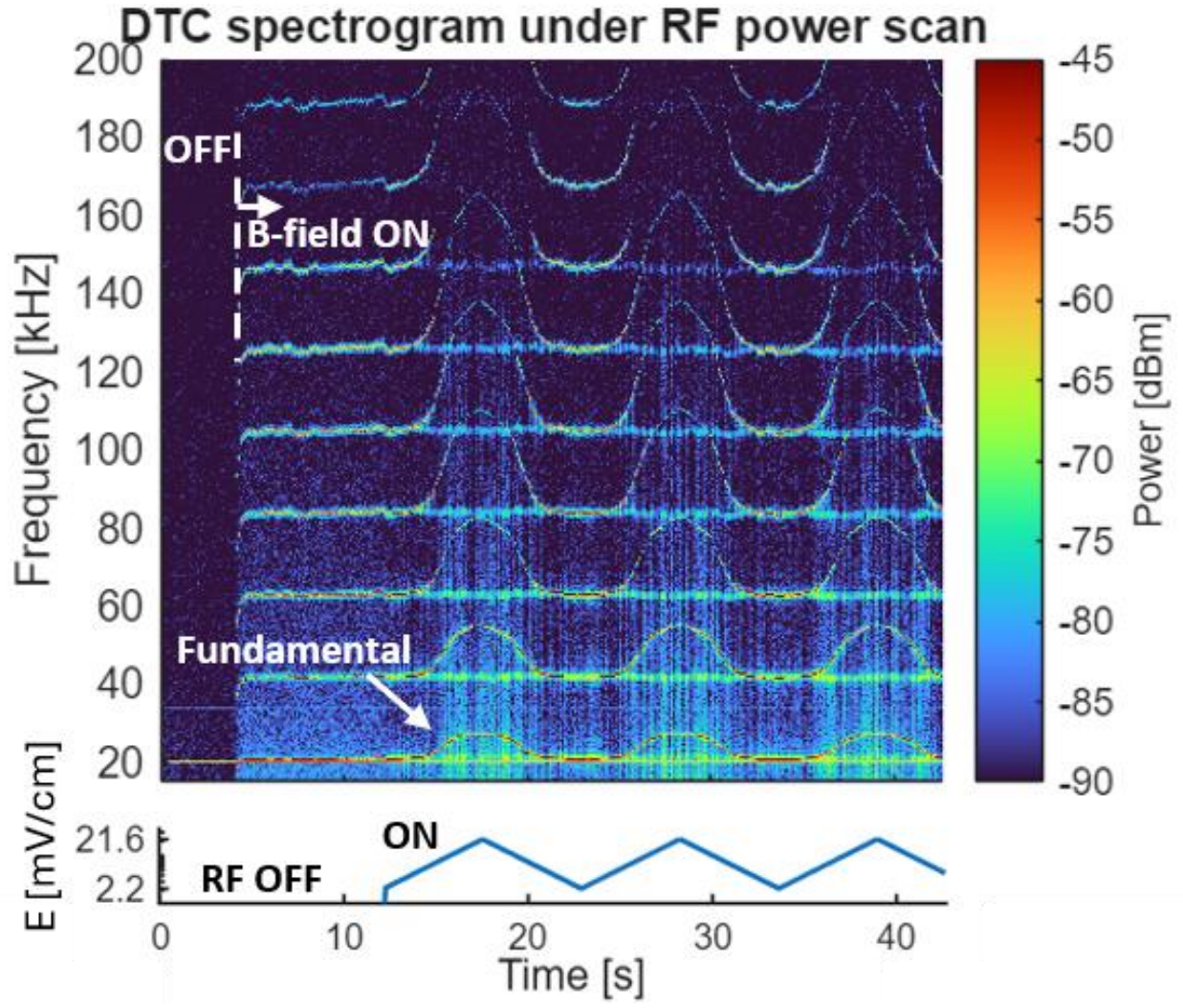

**Figure 4. RF-driven DTC frequency evolution.** Spectrogram of the photo-detected Rydberg-EIT signal with the lasers locked and a static magnetic field of approximately 14 G applied to establish the DTC oscillation. The RF field is initially off, then swept repeatedly between an estimated $E_{rms} \approx 2.2$ and 21.6 mV/cm using a triangular source-power waveform. The fundamental DTC oscillation near 21 kHz shifts systematically with RF field, while higher harmonics undergo correspondingly larger absolute frequency excursions. The lower trace shows the applied RF-field waveform versus time on a logarithmic field axis; the apparent triangular form reflects the source sweep being linear in dBm.

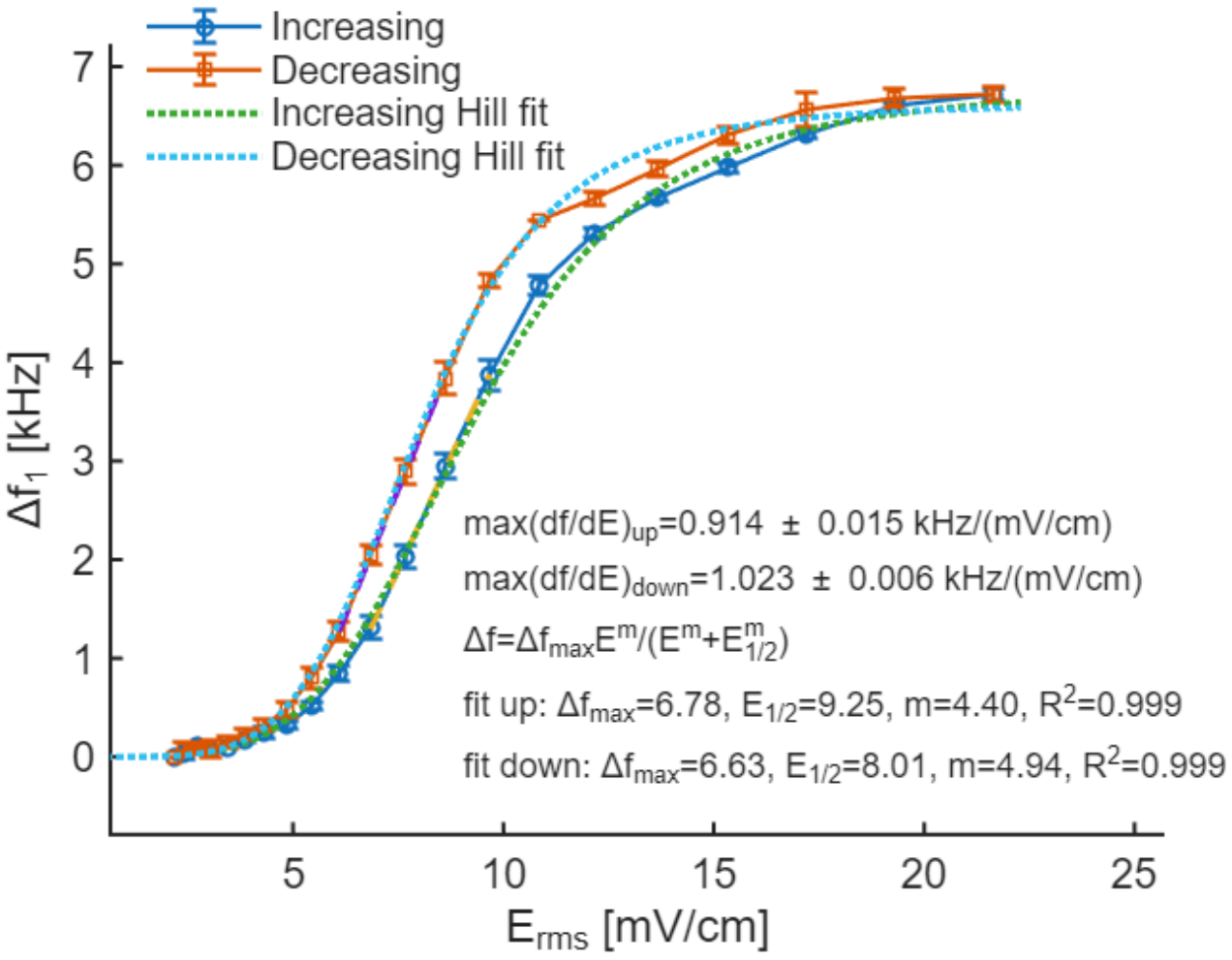

**Figure 5. RF field-to-frequency transduction of the magnetically initiated Rydberg DTC.** Fundamental DTC frequency shift, $\Delta f_1$, versus estimated rms RF field for increasing and decreasing power sweeps; error bars show cycle-to-cycle standard deviation. The response transitions from a weak-field plateau to a high-responsivity region near $8-10$ mV/cm and saturates near 6.7 kHz. Maximum local responsivities are $0.914 \pm 0.015$ and $1.023 \pm 0.006$ kHz/(mV/cm) for increasing and decreasing sweeps, respectively. Dotted curves are fits to $\Delta f = \Delta f_{max} E^m / (E^m + E^m_{1/2})$, yielding $m = 4.40$ and 4.94 with $R^2 = 0.999$. The modest shift between sweep directions indicates a small dynamic hysteresis/lag in the RF-driven DTC response.

The resulting field-to-frequency transfer characteristic is strongly nonlinear (Fig. 5). The fundamental frequency is only weakly perturbed at low $E_{rms}$, followed by a narrow region of rapidly increasing response centered near $8–10\ \mathrm{mV/cm}$, before approaching a total shift of approximately 6.7kHz at higher field. The largest measured local slopes are $0.914 \pm 0.015$ and $1.023 \pm 0.006\ \mathrm{kHz/(mV/cm)}$ for increasing and decreasing sweeps, respectively, corresponding to approximately a 1-kHz change in DTC frequency for a $1\ \mathrm{mV/cm}$ field variation near the point of maximum responsivity. The two sweep directions retain nearly the same overall transfer characteristic but exhibit a small displacement of the transition region, consistent with weak dynamical lag or hysteresis rather than abrupt switching between distinct DTC branches. Phenomenological Hill fits reproduce the full response with $R^2 = 0.999$, yielding nonlinear-transition midpoints of 9.25 and 8.01 mV/cm and effective exponents $m = 4.40$ and 4.94 for increasing and decreasing field, respectively. These large effective exponents quantify a crossover substantially sharper than a simple quadratic saturation response and are consistent with collective-feedback dynamics.

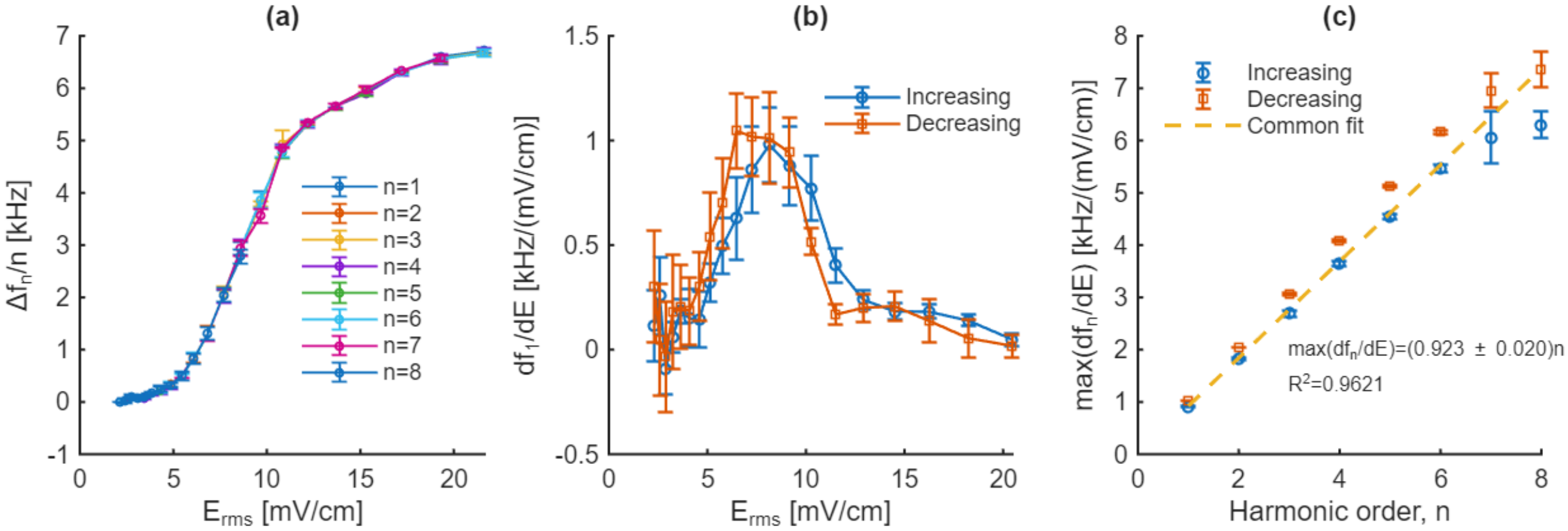

**Figure 6. Harmonic scaling and differential RF responsivity of the Rydberg DTC.** (a) Harmonic frequency shifts normalized by harmonic order, $\Delta f_n/n$, collapse onto a common field-dependent response for $n = 1$–$8$, showing that the observed spectral branches are harmonics of the same underlying DTC oscillation. (b) Local fundamental responsivity, $df_1/dE$, obtained directly from adjacent measured field points for increasing and decreasing RF sweeps. The responsivity peaks near the nonlinear transition region and decreases as the frequency response approaches saturation. (c) Maximum local responsivity, $\max(df_n/dE)$, versus harmonic order. The approximately linear scaling with $n$ is captured by a common through-origin fit, $\max(df_n/dE) = (0.923 \pm 0.020)n$ kHz/(mV/cm), with $R^2 = 0.962$, demonstrating progressively larger absolute frequency transduction at higher DTC harmonics.

The enhanced steepness can be described by a minimal self-consistent mean-field model. Let $x$ denote an effective RF-induced redistribution of the participating Rydberg population. For a near-resonantly driven Rydberg transition, we take $x = s/(1+s)$, with $s = \alpha E^2/[1 + (\Delta_{\text{eff}}/\Gamma_2)^2]$, where $\Gamma_2$ is an effective coherence width and $\alpha$ contains the transition dipole moment and relaxation parameters. Established mean-field descriptions of interacting Rydberg gases and DTCs include population-dependent collective resonance shifts [10, 13]. We therefore write $\Delta_{\text{eff}} = \Delta_0 - \kappa x$, where $\kappa$ is an effective nonlinear coupling coefficient relating Rydberg-population redistribution to the collective resonance shift. This coefficient may incorporate direct or microwave-dressed Rydberg interactions and possible charge-induced Stark contributions. When the population-induced shift moves the microwave-coupled manifold toward resonance, the feedback $E \to x \to \Delta_{\text{eff}} \to x$ enhances the field dependence beyond the bare quadratic response.

The enhancement can be quantified by the local field exponent $n_{\text{eff}} \equiv d\ln x/d\ln E$, yielding

$$n_{\text{eff}} = \frac{2}{1 + s - \dfrac{2\kappa\Delta_{\text{eff}}x}{\Gamma_2^2 + \Delta_{\text{eff}}^2}}.$$

Without collective feedback ($\kappa = 0$), $n_{\text{eff}} = 2/(1+s) \le 2$, so ordinary saturation only softens the underlying $E^2$ dependence. With $\Delta_{\text{eff}} = \Delta_0 - \kappa x$, the feedback is enhanced when $\kappa\Delta_{\text{eff}} > 0$ such that increasing Rydberg population moves the effective transition toward resonance. A locally superquadratic response occurs when the resonance-enhancement term exceeds the saturation contribution, $2\kappa\Delta_{\text{eff}}x/(\Gamma_2^2 + \Delta_{\text{eff}}^2) > s$, giving $n_{\text{eff}} > 2$. This provides a physical basis for the large phenomenological Hill exponents $m \approx 4 - 5$ without requiring a higher-order microscopic microwave interaction. The model is intended to explain the enhanced field dependence rather than the full DTC limit cycle or absolute oscillation frequency. It also predicts that the transduction slope should depend on operating detuning; future work should investigate the feedback picture and distinguish interaction-mediation from Stark-shift contributions.

## III. HARMONIC SCALING OF RF RESPONSIVITY

The higher-order spectral components provide an additional test of whether the microwave response originates from a common DTC limit cycle. As shown in Fig. 6(a), normalizing the measured frequency shift by harmonic order, $\Delta f_n/n$, collapses the responses for $n = 1$–$8$ onto a common field-dependent curve within experimental uncertainty. Thus, the spectral branches satisfy approximately $\Delta f_n(E) = n\Delta f_1(E)$, confirming that they represent harmonics of the same RF-tuned DTC oscillation rather than independent microwave-induced modes. The differential fundamental response, $df_1/dE$, in Fig. 6(b) further emphasizes the nonlinear character of the transduction: the responsivity is small at low field, rises sharply through the transition region to approximately $1\ \text{kHz}/(\text{mV/cm})$, and decreases as the frequency shift approaches saturation.

Because $f_n = nf_1$, the absolute frequency responsivity is expected to scale proportionally with harmonic order. Figure 6(c) confirms this behavior experimentally, with the maximum local responsivity following $\max(df_n/dE) = (0.923 \pm 0.020)n\,\mathrm{kHz}/(\mathrm{mV/cm})$ and $R^2 = 0.962$. Higher DTC harmonics therefore provide proportionally larger absolute frequency excursions for a given RF-field change and may improve field resolution when the harmonic frequency uncertainty does not scale proportionally with harmonic order. More generally, the harmonic spectrum provides multiple frequency-domain observables of the same underlying RF transduction response, which could be combined for improved field estimation.

## IV.CONCLUSION

Near-resonant microwave/RF power to DTC-frequency transduction is demonstrated in a magnetically initiated Rydberg dissipative time crystal. Microwave coupling of the $63D_{3/2} \rightarrow 64P_{3/2}$ manifold continuously shifts the autonomous DTC frequency, producing a strongly nonlinear field-to-frequency transfer function with peak responsivity near $1\,\mathrm{kHz}/(\mathrm{mV/cm})$ and a total fundamental-frequency excursion of approximately 6.7 kHz. A minimal self-consistent mean-field model shows that population-dependent collective resonance shifts can enhance the underlying quadratic microwave response, providing a physical basis for the observed superquadratic crossover without requiring a higher-order microscopic RF interaction. Higher DTC harmonics preserve the same underlying transfer function while providing proportionally larger absolute frequency excursions.

The resulting architecture differs from conventional spectroscopic and heterodyne Rydberg receivers by encoding RF amplitude directly into an emergent atomic DTC frequency that can be tracked continuously at a fixed optical operating point. The power-to-frequency conversion arises from the nonlinear atomic dynamics themselves rather than from an externally engineered RF feedback oscillator such as in [6], while avoiding repeated optical spectral scans or resolved Autler–Townes splitting. This establishes a distinct many-body route to frequency-encoded RF sensing. Determination of the ultimate field sensitivity will require measurements of DTC frequency noise, its dependence on microwave and optical detuning, and the noise scaling of the higher harmonics.

## ACKNOWLEDGMENTS

The research was carried out at the Jet Propulsion Laboratory, California Institute of Technology, under a contract with the National Aeronautics and Space Administration (80NM0018D0004), through the Instrument Incubator Program's (IIP) Instrument Concept Development (Task Order 80NM0022F0020).

## AUTHOR CONTRIBUTIONS

D.A. proposed the project. D.A. configured the atomic systems to include lasers, and stabilization/locking systems. D.A. developed the digital system for locking and laser error reduction. D.A. developed and optimized excitation and field sources for DTC structure observation. D.A. designed the software scripts for data collection and processed the data for the figures. D.A. conducted all data collection efforts. D.A prepared the manuscript.

## DATA AVAILABILITY STATEMENT

The data that support the findings of this study are available from the corresponding author, Darmindra Arumugam (email: darmindra.d.arumugam@jpl.nasa.gov).

[1]Sedlacek, J., Schwettmann, A., Kübler, H. et al. Microwave electrometry with Rydberg atoms in a vapour cell using bright atomic resonances. Nature Phys 8, 819–824 (2012).
[2]C. L. Holloway, J. A. Gordon, S. Jefferts, A. Schwarzkopf, D. A. Anderson, S. A. Miller, N. Thaicharoen, and G. Raithel, "Broadband Rydberg atom-based electric-field probe for SI-traceable, self-calibrated measurements," *IEEE Trans. Antennas Propag.* **62**, 6169–6182 (2014).
[3]M. Jing, Y. Hu, J. Ma, H. Zhang, L. Zhang, L. Xiao, and S. Jia, "Atomic superheterodyne receiver based on microwave-dressed Rydberg spectroscopy," *Nat. Phys.* **16**, 911–915 (2020).
[4]S. Middelhoek, P. J. French, J. H. Huijsing, and W. J. Lian, "Sensors with digital or frequency output," *Sens. Actuators* **15**, 119–133 (1988).
[5]T. Dinh, M. Rais-Zadeh, T. Nguyen, H.-P. Phan, P. Song, R. Deo, D. Dao, N.-T. Nguyen, and J. Bell, "Micromachined mechanical resonant sensors: From materials, structural designs to applications," *Adv. Mater. Technol.* **9**, 2300913 (2024).
[6]D. Arumugam, "Microwave power-to-frequency transduction via injection pulling of a self-sustained oscillator for Rydberg superheterodyne sensing," arXiv:2605.08535 (2026).
[7]C. Carr, R. Ritter, C. G. Wade, C. S. Adams, and K. J. Weatherill, "Nonequilibrium phase transition in a dilute Rydberg ensemble," *Phys. Rev. Lett.* **111**, 113901 (2013).
[8]D.-S. Ding, Z.-K. Liu, B.-S. Shi, G.-C. Guo, K. Mølmer, and C. S. Adams, "Enhanced metrology at the critical point of a many-body Rydberg atomic system," *Nat. Phys.* **18**, 1447–1452 (2022).
[9]Z. Zhang, Z. Zhang, S. Han, Y. Zhang, G. Zhang, J. Wu, V. B. Sovkov, W. Liu, Y. Li, L. Zhang, L. Xiao, S. Jia, W. Li, and J. Ma, "Microwave-coupled optical bistability in driven and interacting Rydberg gases," *npj Quantum Inf.* **11**, 44 (2025).
[10]X. Wu, Z. Wang, F. Yang, R. Gao, C. Liang, M. K. Tey, X. Li, T. Pohl, and L. You, "Dissipative time crystal in a strongly interacting Rydberg gas," *Nat. Phys.* **20**, 1389–1394 (2024).
[11]B. Liu, L.-H. Zhang, Y. Ma, Q.-F. Wang, T.-Y. Han, J. Zhang, Z.-Y. Zhang, S.-Y. Shao, Q. Li, H.-C. Chen, and G.-C. Guo, "Bifurcation of time crystals in driven and dissipative Rydberg atomic gas," *Nat. Commun.* **16**, 1419 (2025).
[12]Y. Jiao, W. Jiang, Y. Zhang, J. Bai, Y. He, H. Shen, J. Zhao, and S. Jia, "Observation of multiple time crystals in a driven-dissipative system with Rydberg gas," *Nat. Commun.* **16**, 8767 (2025).
[13]D. Arumugam, "Electric-field sensing with driven-dissipative time crystals in room-temperature Rydberg vapor," *Sci. Rep.* **15**, 13446 (2025).
[14]D. Arumugam, "Stark modulated Rydberg dissipative time crystals at room temperature applied to sub-kHz electric field sensing," *Sci. Rep.* **15**, 35976 (2025).
[15]D. Arumugam, "Injection locking of Rydberg dissipative time crystals," *Commun. Phys.* **9**, 156 (2026).
[16]D. Manchaiah, W. J. Watterson, and C. L. Holloway, "Frequency comb behavior of time crystals in an rf-driven dissipative Rydberg system," *Phys. Rev. Res.* **8**, 033115 (2026).